\pdfoutput=1

\newcommand{\schedulerbase}{\gls{bfd}}
\newcommand{\schedulerfreq}{\gls{bcffs}}
\newcommand{\schedulermigr}{\gls{bcf}}

\newcommand{\ensavingsmax}{$32\%$}
\newcommand{\ensavingsfreq}{$8\%$}
\newcommand{\vmnumsimulation}{2k}
\newcommand{\pmnumsimulation}{2k}
\newcommand{\maxactivepmnumsimulation}{1k}

\documentclass[conference]{IEEEtran}

\usepackage[T1]{fontenc} \usepackage[utf8]{inputenc}
\usepackage{url}
\usepackage{graphicx}
\DeclareGraphicsExtensions{.pdf,.png,.jpg}
\graphicspath{{./img/}}
\usepackage{algpseudocode}
\usepackage{flushend}
\usepackage{amsmath}
\usepackage{float}
\usepackage{amssymb}
\usepackage{glossaries}
\usepackage[super]{nth}
\usepackage{mathptmx}
\usepackage{todonotes}
\newcommand{\coe}{$CO_{2}e$}

\newcommand{\tabletopmargin}{-0cm}
\newcommand{\tablecaptionmargin}{-0cm}
\newcommand{\tablebottommargin}{-0cm}
\newcommand{\figtopmargin}{-0cm}
\newcommand{\figbottommargin}{-0cm}
\newcommand{\figcaptionmargin}{-0cm}

\newacronym{coe}{\coe}{$CO_{2}$ equivalent}
\newacronym{ewma}{EWMA}{exponentially weighted moving average}
\newacronym{dvfs}{DVFS}{dynamic voltage \& frequency scaling}
\newacronym{vm}{VM}{virtual machine}
\newacronym{era}{ERA}{energy reduction assets}
\newacronym{api}{API}{application programming interface}
\newacronym{os}{OS}{operating system}
\newacronym{rtp}{RTP}{real-time pricing}
\newacronym{qos}{QoS}{quality of service}
\newacronym{sla}{SLA}{service level agreement}
\newacronym{rtep}{RTEP}{real-time electricity pricing}
\newacronym{iaas}{IaaS}{infrastructure as a service}
\newacronym{pm}{PM}{physical machine}
\newacronym{pue}{PUE}{power usage efficiency}
\newacronym{cue}{CUE}{carbon usage effectiveness}
\newacronym{cef}{CEF}{carbon emission factor}
\newacronym{hpc}{HPC}{high-performance computing}
\newacronym{db}{DB}{database}
\newacronym{dc}{DC}{data center}
\newacronym{oltp}{OLTP}{online transaction processing}
\newacronym{mse}{MSE}{mean squared error}
\newacronym{ga}{GA}{genetic algorithm}
\newacronym{arima}{ARIMA}{autoregressive integrated moving average}
\newacronym{ses}{SES}{simple exponential smoothing}
\newacronym{bcf}{BCF}{best cost fit}
\newacronym{bfd}{BFD}{best fit decreasing}
\newacronym{iot}{IoT}{Internet of Things}
\newacronym{bcffs}{BCFFS}{Best Cost Fit Frequency Scaling}
\newacronym{fps}{FPS}{frames per second}
\newacronym{hevc}{HEVC}{high efficiency video coding}

\usepackage{tikz}

\usetikzlibrary{fit,positioning}
\usetikzlibrary{arrows}
\usetikzlibrary{calc}

\usepackage{booktabs}
\usepackage{algpseudocode}
\usepackage{algorithm}
\usepackage{varwidth}
\usepackage{relsize}

\begin{document}

\title{A Cloud Controller for Performance-Based Pricing}

\author{
	\IEEEauthorblockN{Dražen Lučanin\IEEEauthorrefmark{1},
	                  Ilia Pietri\IEEEauthorrefmark{2},
					  Ivona Brandic\IEEEauthorrefmark{1},
	                  Rizos Sakellariou\IEEEauthorrefmark{2}
	}
	\IEEEauthorblockA{
		\IEEEauthorrefmark{1}
		Vienna University of Technology, Vienna, Austria\\
 		Email: drazen.lucanin@tuwien.ac.at, ivona@ec.tuwien.ac.at
 	}
 	\IEEEauthorblockA{
		\IEEEauthorrefmark{2}
		University of Manchester, Manchester, UK\\
 		Email: \{pietrii, rizos\}@cs.man.ac.uk
 	}
}

\maketitle 

\begin{abstract}
New dynamic cloud pricing options are emerging with cloud providers offering resources as a wide range of CPU frequencies and matching prices that can be switched at runtime. On the other hand, cloud providers are facing the problem of growing operational energy costs. This raises a trade-off problem between energy savings and revenue loss when performing actions such as CPU frequency scaling. Although existing cloud controllers for managing cloud resources deploy frequency scaling, they only consider fixed \gls{vm} pricing. In this paper we propose a performance-based pricing model adapted for \gls{vm}s with different CPU-boundedness properties. We present a cloud controller that scales CPU frequencies to achieve energy cost savings that exceed service revenue losses. We evaluate the approach in a simulation based on real \gls{vm} workload, electricity price and temperature traces, estimating energy cost savings up to \ensavingsmax{} in certain scenarios.
\end{abstract}

\section{Introduction}


With the wide range of \gls{vm} types, heterogeneous infrastructure and different computing environments including \gls{vm}s and containers~\cite{felter_updated_2014}, estimating the performance of provisioned resources is becoming increasingly challenging~\cite{oloughlin_performance_2014}.\
New cloud pricing schemes are emerging where resources are priced based on the delivered performance.\
%
%
For example, the CPU frequency provided to a \gls{vm} at runtime determines the price \cite{pietri2014cost},\
with high CPU frequencies being more expensive.\
We call this model performance-based pricing and it is used in production by cloud providers, such as ElasticHosts~\cite{elastichosts}.\
Though this approach mainly targets users, it could also be used by cloud providers to control their energy consumption. Energy consumption of data centers is becoming a major issue, accounting for 1.5\% of global electricity usage \cite{jonathan_koomey_growth_2011}. Furthermore, modern clouds may consist of geographically-distributed data centers influenced by dynamic local factors, such as real-time electricity prices \cite{weron_modeling_2006} and temperature-dependent cooling \cite{xu_temperature_2013}, that we call geotemporal inputs.\

We call the subsystem of the cloud, that determines the\
actions\
to allocate and manage the \gls{vm}s, a \emph{cloud controller}.\
Adapting the cloud controller to geotemporal inputs through actions,
such as CPU frequency scaling, raises a trade-off problem between the potential energy savings and service revenue losses incurred under performance-based pricing.\
This is the challenge at the core of this paper.\


Frequency scaling is a power management technique commonly used to lower the operating frequency of hardware resources in order to reduce power consumption\
\cite{miyoshi2002critical}.\
However, frequency reduction may degrade\ 
the performance of resources.\ 
Depending on the workload characteristics, workload performance may be affected in different ways by the resource's operating frequency~\cite{freeh2007analyzing}. E.g., CPU-bound workloads are more sensitive to the provided CPU frequency.\
On the other hand, the performance of I/O-bound workloads is less sensitive to frequency reduction.\
As the operating frequency may affect workload performance, pricing models can be used as a mechanism to offer motivation to users for configurations with different speed and cost characteristics, with the price of each \gls{vm} adjusted based on the perceived performance level.\
For example, users of services with heterogeneous hardware,\
such as Amazon EC2,\ 
would benefit from a pricing scheme that takes into account the volatile hardware performance by being charged based on the performance perceived by the \gls{vm}~\cite{oloughlin_performance_2014}.\ 
%
%
The gross profit from energy savings and service revenue losses may not be positive for some CPU frequency scales.
Pricing schemes can be used by the providers to find configurations where the energy cost savings exceed the service revenue losses\
to balance the trade-offs.\

Existing work on CPU frequency scaling and \gls{vm} migration aims to reduce energy consumption without significant impact on workload performance \cite{von2009power,shi2011towards}. However, such work is limited to fixed pricing and does not consider\
performance-based pricing.\
Also, geotemporal inputs of data centers, including electricity prices and temperature-based cooling, are not explored in existing methods when deploying \gls{vm} migration \cite{beloglazov_energy-aware_2012} or CPU frequency scaling actions \cite{wu2014green}. Methods that consider geotemporal inputs \cite{guler_cutting_2013,liu_renewable_2012} only perform initial job placement, without considering reallocation through \gls{vm} migration or frequency scaling.


In this paper, we present a \emph{novel cloud controller} suitable for performance-based pricing\
that is invoked periodically\
(e.g. on new \gls{vm} requests or geotemporal input changes)\
to apply \gls{vm} migrations and CPU frequency scaling.\
We firstly develop a pricing model that can be applied for energy-aware cloud control based on the actual impact that CPU frequency scaling will have on a \gls{vm}'s performance.
This means that the price is determined by the performance perceived by the \gls{vm} user based on the workload characteristics, as opposed to the performance provided to it. Hence, we call this model \emph{perceived-performance pricing}.\ 
The model\ 
computes the \gls{vm} price based on the CPU frequency and according to the CPU-boundedness of each \gls{vm}.\ 
Secondly, we propose\
\schedulerfreq{}, a cloud controller we developed.\
The idea behind this cloud controller is to combine frequency scaling and \gls{vm} migrations to reduce energy costs for \gls{vm}s as long as the performance-based revenue losses do not exceed the energy cost savings.\
It is a two-stage algorithm that first migrates \gls{vm}s based on geotemporal inputs and in the second stage applies CPU frequency scaling based on the energy-revenue trade-off problem.

We evaluate the \schedulerfreq{} cloud controller by comparing it to two baseline controllers \cite{beloglazov_energy-aware_2012} in a trace-based simulation using the Philharmonic framework we developed \cite{lucanin2014energy}.\
We compute the service revenue and energy cost\ 
based on historical traces of real-time electricity prices~\cite{alfeld_toward_2012} and temperatures.\ 
CPU-boundedness values in the simulation are distributed according to the PlanetLab\ 
dataset of \gls{vm} CPU usage. We show that energy savings up to \ensavingsmax{} without significant service revenue reductions are possible using the \schedulerfreq{} cloud controller.

The \textbf{key contributions} are: (1) We develop a perceived-performance pricing model for determining the \gls{vm} price based on the provided CPU frequency and workload CPU-boundedness. (2) We propose a \schedulerfreq{} cloud controller for \gls{vm} migration and CPU frequency scaling by balancing the trade-offs of service revenue loss and energy savings. (3) We evaluate the controller in a simulation based on realistic CPU-boundedness and geotemporal input traces, providing insights into parameters important for the efficiency of the controller.

After examining the related work in the next section, in Section~\ref{sec:problem} we describe the considered problem. In Section~\ref{sec:model} we present the power and pricing models. In Section~\ref{sec:scheduler} we explain our \schedulerfreq{} cloud controller. In Section~\ref{sec:evaluation} we present the evaluation methodology and\ 
the most significant results. We provide our concluding remarks in Section~\ref{sec:conclusion}.

\section{Related Work}
\label{sec:related}


Adapting distributed systems to geotemporal inputs has been studied previously. In \cite{qureshi_cutting_2009}, network routing in content delivery networks is adapted for \gls{rtep}. Savings of up to 40\% of the full electricity cost are estimated.\ 
Job placement based on geotemporal inputs for map-reduce jobs is researched in \cite{buchbinder_online_2011} and for computational grids based on both \gls{rtep} and cooling in \cite{guler_cutting_2013,liu_renewable_2012}.\
However, geotemporal inputs as a basis for scaling CPU frequencies or as a counter-balance to performance-based pricing has not been researched. 

A lot of work focuses on power management techniques and particularly frequency scaling \cite{von2009power,wu2014green,etinski2010optimizing}. \ 
In \cite{wu2014green}, a cloud scheduler to prioritise and allocate jobs to \gls{vm}s taking into account the required \gls{sla} of the users is proposed. The algorithm reduces energy consumption by allocating the minimum resource requirements to \gls{vm}s in order to avoid resource wastage and controlling the operating frequencies of the hosts under light workloads, without degrading the performance of the executing jobs. Job scheduling to \gls{vm}s using frequency scaling is also the subject in \cite{von2009power}.\
The proposed approach scales the frequencies of the cluster servers at runtime and schedules the queued \gls{vm}s to servers where \gls{vm} performance requirements can be met. The decision is based on their power profiles, preferring servers that operate at lower frequencies. In contrast to related work, our adaptive approach scales the operating frequencies based on the CPU-boundedness of the mapped \gls{vm}s while assessing the impact frequency reduction has on the provider's gross profit.

Also, the impact of frequency scaling on the system and workload performance is investigated in many studies \cite{miyoshi2002critical,freeh2007analyzing,etinski2010optimizing,hsu2003design}.\ 
The proposed compiler algorithm in \cite{hsu2003design} aims at identifying program regions with low CPU utilization to scale the operating frequency and reduce energy consumption without impacting workload performance significantly. The authors introduce a metric to model workload CPU-boundedness, based on the idea that potential energy savings from frequency scaling depend on the CPU-boundedness of the benchmark, with more CPU-bound workloads having lower energy savings. The work in \cite{miyoshi2002critical} investigates the power-performance features of systems that support power management techniques, introducing the critical power slope concept to determine the operating performance points of a system that lead to energy savings. Metrics to predict the energy-performance trade-off and determine the operating gears to use are also the focus in \cite{freeh2007analyzing}, investigating the impact of frequency scaling on workload performance for different HPC workloads.

\section{Problem Description}
\label{sec:problem}

In this work, we consider a single \gls{iaas} cloud provider with a large number of physical machines (or hosts) located among a set of geographically distributed data centers. The cloud is influenced by geotemporal inputs, such as real-time electricity prices and temperature-dependent cooling efficiency, which are different for each data center. Each \gls{pm} has a specified number of resource types\
with maximum capacities (e.g. CPU core number or RAM amount)\
and can operate at a number of available CPU frequencies.\

Each \gls{pm} can host multiple \gls{vm}s to serve the users' requests.\ 
For example, a user can request a \gls{vm} with 1 CPU core and 1024 MB of RAM.\ 
Depending on the host's operating frequency and the \gls{vm}'s CPU usage characteristics, the performance level of the \gls{vm} may vary. For example, performance degradation that results from CPU frequency reduction is lower in the case of less CPU-bound workloads. A parameter ($\beta$) to characterize the CPU-boundedness of a workload based on the computation time at the corresponding operating frequencies is introduced in \cite{hsu2003design} and investigated in other studies \cite{freeh2007analyzing,etinski2010optimizing}. The parameter ranges between 0 and 1, with more I/O-bound \gls{vm}s taking values close to 0 and CPU-bound \gls{vm}s close to 1.\
In our proposed model, the parameter $\beta$ is used as a performance metric to characterise the sensitivity of the \gls{vm} performance on frequency scaling based on the CPU usage. Each \gls{pm} can access the CPU usage of the \gls{vm}s and monitor the impact of CPU on \gls{vm} performance \cite{mastelic2014novel}. 


The virtual and physical machines are managed by a cloud controller which determines the actions to be performed. These include the migration of a \gls{vm} to a \gls{pm}, suspending or resuming a \gls{pm} and the increase or decrease of the operating frequency of a \gls{pm}. \gls{vm} migration actions result in a migration cost overhead modelled in \cite{liu_performance_2011}, while the transition overhead from frequency scaling is considered to be negligible \cite{etinski2010optimizing}, as well as suspending or resuming empty \gls{pm}s \cite{meisner_powernap:_2009}.

\begin{figure}
\vspace{\figtopmargin}
\centering
\includegraphics[width=1.0\columnwidth]{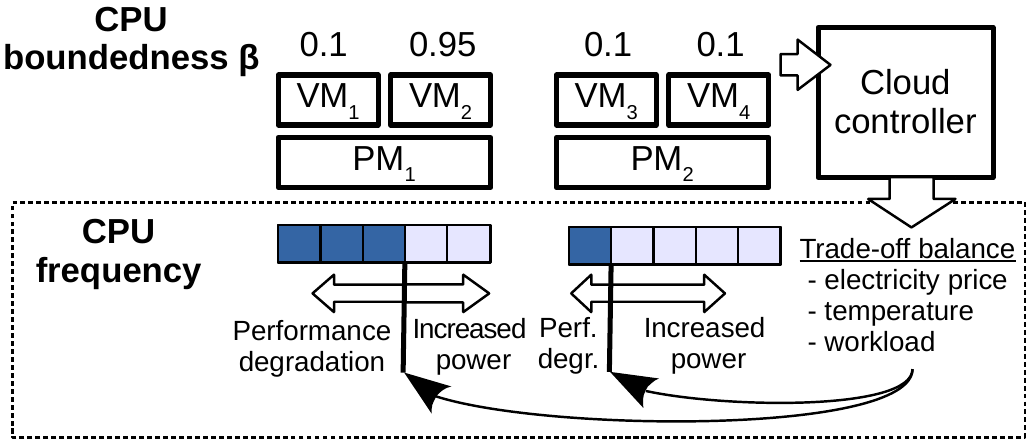}
\vspace{\figcaptionmargin}
\caption{Motivational example.}
\label{fig:motivation}
\vspace{\figbottommargin}
\vspace{-0.4cm}
\end{figure}

To illustrate the problem,\ 
we provide a motivational example in Fig.~\ref{fig:motivation} with two \gls{pm}s hosting \gls{vm}s with different CPU-boundedness properties.\ 
The workload consisting of \gls{vm}s with their $\beta$ values is given as input to the cloud controller,\
which determines the CPU frequency scaling actions\
to apply to the \gls{pm}s as the output.\
The first machine, $PM_1$, hosts $VM_1$ with low CPU-boundedness and $VM_2$ with high CPU-boundedness, while the second physical machine, $PM_2$, hosts $VM_3$ and $VM_4$, both with low CPU-boundedness.\
If the cloud operates all hosts at a maximum frequency, the allocated \gls{vm}s receive good performance, however an increased power consumption may lead to power wastage for \gls{vm}s 1, 3 and 4 that are not CPU-bound. In order to reduce power consumption, the operating frequencies of the hosts could be adjusted to a minimum frequency. Although this scenario leads to energy savings, it also leads to performance degradation of the CPU-bound $VM_2$.\ 

A desirable strategy would be to analyse the trade-offs between energy savings and performance to adaptively select an optimal frequency for each \gls{pm} independently as shown in Fig.~\ref{fig:motivation}.\
The balance between the two goals depends on many factors such as the workload characteristics, electricity prices and temperatures which all have to be taken into account to determine the optimal frequency.\ 
In this ideal scenario, the operating frequencies of \gls{pm}s hosting \gls{vm}s with low CPU-boundedness are reduced, e.g. in the case of $PM_2$. On the other hand, \gls{pm}s hosting more CPU-bound \gls{vm}s are kept high, e.g. in the case of $PM_1$, which is set to operate at a high-enough frequency to avoid performance degradation for $VM_2$.

%

The goal of our work is to develop such an adaptive approach 
that takes into account the CPU-boundedness properties of \gls{vm}s to achieve energy cost savings without significantly reducing the performance.\
\
The challenges in achieving this goal are:\
(1) It is necessary to quantitatively determine the balance between energy savings and workload performance to find the optimal CPU frequency.\
(2) When considering geotemporal inputs, changes in electricity prices and temperatures impact the energy cost\ 
and it is necessary to reevaluate the control actions at runtime.
To address these challenges, in the next section we present models\ 
that allow the comparison of energy savings based on geotemporal inputs and the performance impact\
quantified as revenue losses caused by frequency scaling in performance-based pricing.



\section{Cost-related Components}

\label{sec:model}
In this section we define the power and service revenue models\ 
that influence cloud costs based on energy consumption\
and user costs based on \gls{vm} prices.\
These two models are used to formulate the\ 
trade-off problem and to determine the actions to be deployed by the cloud controller.

\subsection{Cloud Cost Model}

Power consumption is modelled based on frequency scaling and CPU utilisation using disparate models. 
The idea is that power consumption of a \gls{pm} increases with higher resource utilisation caused from more \gls{vm}s per \gls{iaas} cloud models and with higher operating CPU frequencies per \gls{hpc} models.\ 
A cubic model based on \cite{pierson2011utility} is used to compute the power consumed when the host is fully utilised according to the operating frequency, $f$. The peak power of the host operating at frequency $f$ is given as: 
\
\begin{equation}\label{P_max_f}
P_{{peak}_f}=P_{base} + P_{dif} (\frac{f-f_{base}}{f_{base}})^3,
\end{equation}
\
where $P_{base}$ is the peak power of the host operating at a minimum frequency $f_{base}$ and $P_{dif}$ is a weight\ 
to compute the power at different frequencies. By combining the model from \cite{liu_renewable_2012} for \gls{pm} power under a certain utilisation $util$ with Eq.~\ref{P_max_f}, we can express an integrated power model as:\ 
\
\begin{equation}\label{p_f}
P_f = P_{idle}+util (P_{{peak}_f} - P_{idle})
\end{equation}
\
where $P_{idle}$ is the power consumed by the \gls{pm} when hosting no \gls{vm}s.\ 
We compute $util$ as a uniformly weighted fraction of the \gls{pm}'s CPU cores and RAM consumed by the hosted \gls{vm}s, as CPU and RAM are shown to be the main contributors of power consumption \cite{basmadjian2011methodology,kansal2010virtual}.\ 
The power model graph is shown in Fig.~\ref{fig:power_model} with \gls{pm}'s power depending on $util$ and CPU frequency.\
We can see that power decreases significantly for even a small reduction of frequency for high \gls{pm} utilisation, due to the cubic shape of the curve. Gradually, the curve becomes less steep for lower utilisation ratios.\
Power consumption of empty \gls{pm}s is considered to be zero, which is approximately possible through fast suspension technology \cite{meisner_powernap:_2009}.

Cooling overhead based on local temperatures\ 
is derived from the power signals of the \gls{pm}s at different data center locations.\
To do so, the model for computer room air conditioning using outside air economisers\
from \cite{xu_temperature_2013} was applied.\
These power signals are then integrated over time and combined with fixed or real-time electricity prices (both models are explored in the evaluation) for the corresponding data center locations to compute the total energy cost of the whole cloud.

\newcommand{\figurepowerpricescale}{0.89}

\begin{figure}[!t]
\vspace{\figtopmargin}
\centering
\includegraphics[width=\figurepowerpricescale\columnwidth]{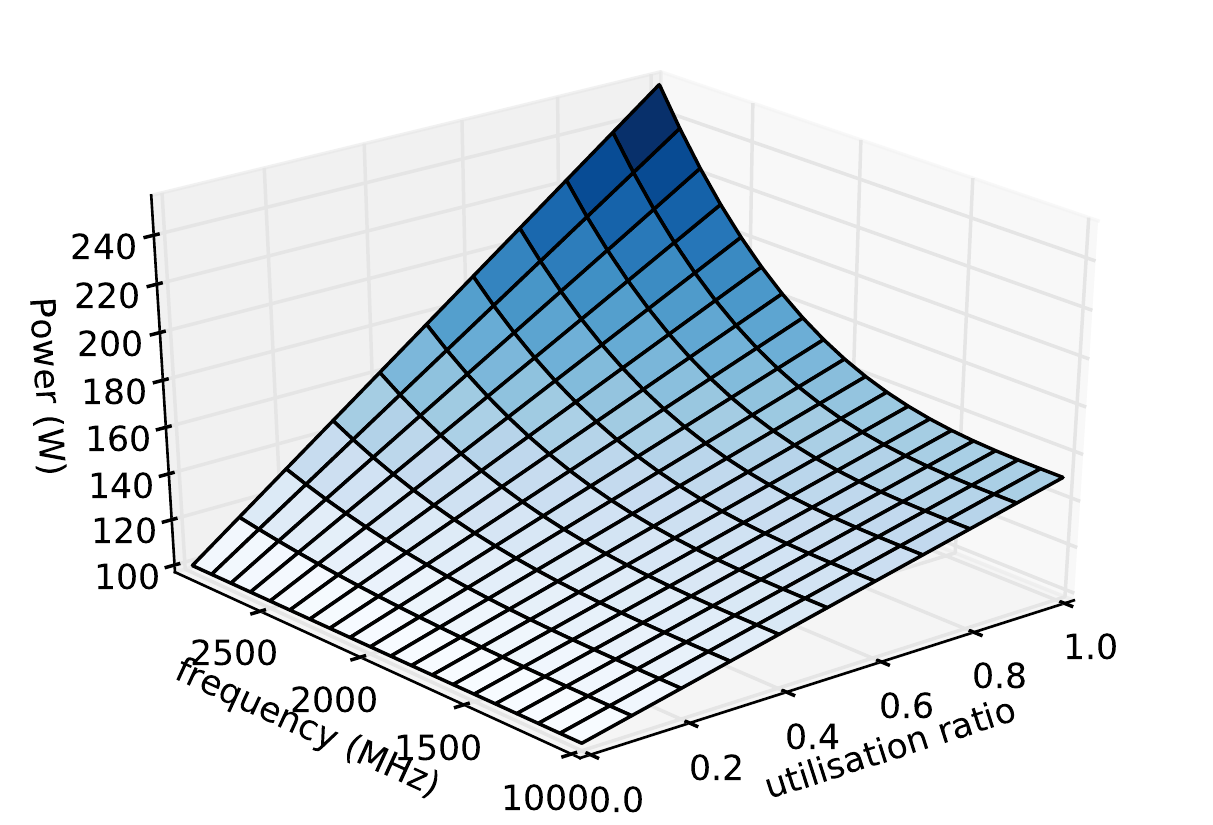}
\caption{Power model.}
\label{fig:power_model}
\vspace{\figbottommargin}
\vspace{-0.3cm}
\end{figure}








\subsection{User Cost Model}

In performance-based pricing, each \gls{vm} is charged\
according to the operating frequency of the \gls{pm} it is mapped to. The model is based on the pricing offered by the ElasticHosts cloud provider~\cite{elastichosts}. The price $C_{vm}$ of each \gls{vm} at frequency $f_{CPU}$ is computed as: 
\
\begin{equation}\label{eq:simple_price}
C_{vm} = C_{base} + C_{CPU} (\frac{f_{CPU}-f_{base}}{f_{base}}) + C_{RAM} (\frac{RAMsize}{RAMsize_{base}}), 
\end{equation}
\
where $C_{base}$ is the \gls{vm} price\ 
at a minimum CPU frequency ($f_{base}$) and RAM size ($RAMsize_{base}$).\ 
$C_{CPU}$ and $C_{RAM}$ are cost weights used to generate the price from the\ 
host's\
CPU frequency $f_{CPU}$ and \gls{vm} RAM size $RAMsize$, respectively. 

\begin{figure}
\vspace{\figtopmargin}
\centering
\includegraphics[width=\figurepowerpricescale\columnwidth]{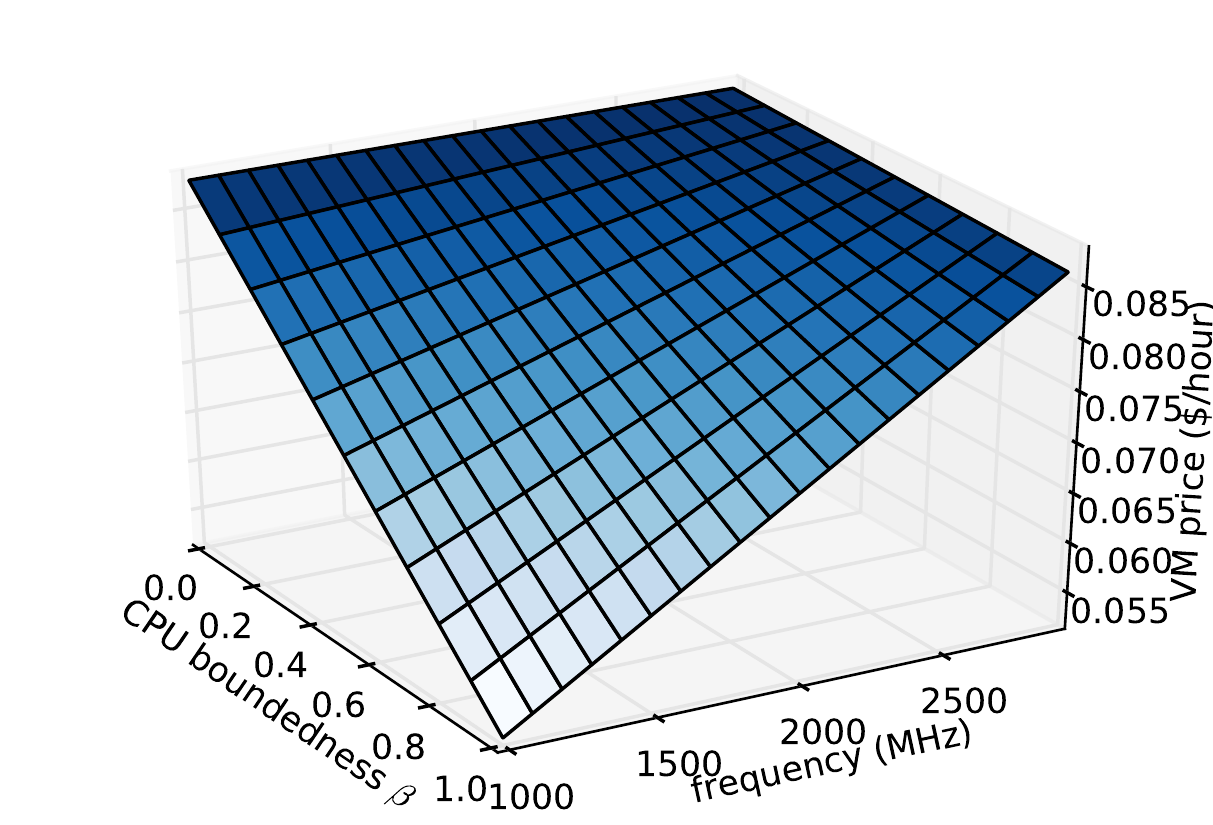}
\caption{Pricing model.}
\label{fig:price_model}
\vspace{\figbottommargin}
\vspace{-0.3cm}
\vspace{-0.1cm}
\end{figure}

Given that the above pricing model does not consider how the \gls{pm}'s CPU frequency affects the \gls{vm}'s performance (as discussed in Section~\ref{sec:problem}), we propose a novel perceived-performance pricing model. In this model, a \gls{vm}'s price is computed based on both the CPU frequency and its impact on workload performance. The idea is that the price may vary between \gls{vm}s where CPU frequency impacts the performance at different degrees. E.g., CPU-bound \gls{vm}s whose performance would be affected by frequency scaling would incur lower monetary costs for the user at a lower frequency than I/O-bound \gls{vm}s. The impact of frequency scaling on workload performance depends on the CPU-boundedness of the job, represented by the parameter $\beta$ \cite{freeh2007analyzing}.\
The pricing model from Eq.~\ref{eq:simple_price} is modified to adjust $f_{CPU}$\ 
to be the frequency perceived by the \gls{vm} computed from the \gls{pm}'s operating frequency $f$, instead of being the \gls{pm}'s CPU frequency directly. We model the perceived frequency $f_{CPU}$ as:\ 
%
\ 
\begin{equation}\label{eq:dyn_price}
f_{CPU}=\beta f + (1 - \beta) f_{max},
\end{equation}
\
where $f_{max}$ is the maximum operating frequency of the host. When a \gls{vm} is CPU-bound, $\beta$ is close to 1 and $f_{CPU}$ changes according to the \gls{pm}'s operating frequency. As a result, the price charged at a lower frequency is also lower, as the user may perceive significant performance degradation. On the other hand, when $\beta$ is close to 0, which corresponds to the scenario of less CPU-bound \gls{vm}s where frequency reduction does not impact performance significantly, $f_{CPU}$ is close to $f_{max}$ and the charged price is less dependent on the \gls{pm}'s operating frequency. A plot of the developed pricing model is shown in Fig.~\ref{fig:price_model}. The axes show the host's operating frequency, the \gls{vm}'s CPU boundedness $\beta$ and\
the resulting \gls{vm} hourly price based on Eq.~\ref{eq:dyn_price}.\ 
We can see that \gls{vm} prices are linearly reduced as a high $\beta$ and a low CPU frequency are approached.

    

\section{Cloud Controller with Frequency Scaling}
\label{sec:scheduler}

With the power and pricing models that allow quantitative comparison of energy saving and revenue loss trade-offs, the next step in addressing the problem from Section~\ref{sec:problem} is\
devising the cloud controller algorithm.\
In this section we present the \schedulerfreq{} cloud controller we developed for \gls{vm} migrations and frequency scaling that increases energy cost savings, as long as they exceed service revenue losses. The controller is invoked periodically, determining the actions to be applied.\
In a production cloud system, the cloud controller would be invoked when new \gls{vm} requests arrive or geotemporal inputs change for a certain threshold, e.g. temperature changes by over 1 C.\
Since the two actions of migrating \gls{vm}s and scaling \gls{pm} CPU frequencies are mutually orthogonal, we examine them separately as two stages of the proposed algorithm.\
In the first stage the controller migrates \gls{vm}s to \gls{pm}s to maximise utilisation and preferring locations with lower electricity and cooling costs. In the second stage, the controller iteratively reduces CPU frequencies of hosts while the resulting energy savings exceed revenue losses. 

\subsection{\gls{vm} Migration Stage}

The first stage\ 
is an algorithm\
that places new \gls{vm}s or reallocates \gls{vm}s from underutilised hosts\
through \gls{vm} migration considering power overhead\
and geotemporal inputs to select hosts.\
Bin packing of \gls{vm}s is an NP-hard problem,\
so we propose a heuristic polynomial time algorithm.

The pseudo-code for this stage of the controller is listed in Alg.~\ref{alg:bcf}.\
The algorithm starts by marking for allocation all newly requested \gls{vm}s\
(line~\ref{alg:bcf:boot}) and for reallocation all \gls{vm}s from underutilised hosts\ 
(line~\ref{alg:bcf:underutil}).\ 
Hosts are defined as underutilised if their utilisation falls bellow a provider-defined threshold, as discussed in \cite{beloglazov_managing_2013}.\
The \gls{vm}s marked for allocation will then be migrated (or initially placed)\
in the outermost loop (line~\ref{alg:bcf:vmloop})\
starting with larger \gls{vm}s first,\
as they are more difficult to fit (line~\ref{alg:bcf:vmsort}).\ 
Available \gls{pm}s are split into $active$ and $nonactive$ lists,\
depending on whether they are suspended or not.\
We sort $inactive$ (line~\ref{alg:bcf:sortinactive})\
so that larger \gls{pm}s come first for activation\
(preferable to more smaller machines, due to the idle power overhead)\
and data centers with lower combined electricity price\
and cooling overhead cost are preferred.\
The target \gls{pm} to host $vm$ is selected in the inner loop\
(line~\ref{alg:bcf:mappedloop}) by sorting $active$\
to try and utilise almost full \gls{pm}s first, then preferring\ 
lower-cost location in case of ties (line~\ref{alg:bcf:sortactive})\
and finally activating the next \gls{pm} from $inactive$ in case $vm$ does not\
fit on any of the $active$ \gls{pm}s (line~\ref{alg:bcf:popinactive}).\
\gls{pm} sorting is the key step of the algorithm, as it assures that data centers will be filled out based on geotemporal inputs.\
When a fitting \gls{pm} is found, the \gls{vm} is placed or migrated to it (line~\ref{alg:bcf:execute})\
and the algorithm continues with the next \gls{vm}.

\begin{algorithm}[!t]
 \scriptsize{
\caption{\gls{vm} Migration Stage.}
\label{alg:bcf}
{
\begin{algorithmic}[1]

\Ensure Allocate or migrate \gls{vm}s per geotemporal inputs.
\Procedure{\gls{vm} Migration Stage}{}
\State $to\_alloc \gets$ empty list
\State append all \gls{vm}s newly requested to $to\_alloc$ \label{alg:bcf:boot} 
\State append \gls{vm}s from all underutilised \gls{pm}s to $to\_alloc$ \label{alg:bcf:underutil}
\State sort $to\_alloc$ by resource requirements decreasing \label{alg:bcf:vmsort}
\For{$vm \in to\_alloc$} \label{alg:bcf:vmloop}
    \State $active \gets $ all PMs where at least one VM is allocated 
    \State $inactive \gets $ all PMs where no VMs are allocated
    \State sort $inactive$ by capacity decreasing, cost increasing\
    \label{alg:bcf:sortinactive}
    \State $mapped \gets False$
    \While{not $mapped$} \label{alg:bcf:mappedloop}
        \State sort $active$ by capacity decreasing, cost increasing \label{alg:bcf:sortactive}
        \For {$pm \in active$}\label{alg:bcf:pmloop}
            \If {$vm$ fits $pm$}\label{alg:bcf:pmfits}
                \State $mapped \gets True$
                \State break loop
            \EndIf
        \EndFor
        \If {not $mapped$}
            \State pop $inactive[0]$ and append it to $active$ \label{alg:bcf:popinactive}
        \EndIf
    \EndWhile
    \State perform a placement/migration of $vm$ to $pm$ \label{alg:bcf:execute}
\EndFor
\EndProcedure
 
\end{algorithmic}
}}
\end{algorithm}

\subsection{Frequency Scaling Stage} 

After determining the actions to be deployed in the initial stage, the algorithm\ 
deploys frequency scaling actions when they lead to energy cost savings that exceed revenue losses based on perceived-performance pricing. 
The polynomial time algorithm is described in Alg.~\ref{alg:frequency_scaling}.\
First, the algorithm resets the frequencies of all the \gls{pm}s to a maximum frequency $f_{max}$ (line~\ref{alg:fs:reset}). Actions do not have to be executed physically until the procedure halts and the final \gls{pm} frequencies are determined.\
Next, the algorithm iterates through the \gls{pm}s in the outer loop (line~\ref{alg:fs:pmiterate}) to find the best CPU frequency for each one.\
The algorithm then iteratively scales down the selected frequency according to the range of available CPU frequencies (line~\ref{alg:fs:freqStep}).

The examined CPU frequency is evaluated based on the following steps:\
The service revenue from the \gls{vm}s hosted by the current \gls{pm} and the energy cost of the \gls{pm} are computed for the previous and the new frequency (lines~\ref{alg:fs:get_revenue}--\ref{alg:fs:get_cost}) based on the power and pricing models presented in Section~\ref{sec:model}. If switching to a new frequency would result in energy cost savings higher than the subsequent revenue loss (line~\ref{alg:fs:comparison}), the new frequency is selected (line~\ref{alg:fs:apply}) and the algorithm continues onto the next lower frequency (line~\ref{alg:fs:freqStep}).\
The inner loop continues until revenue losses surpass energy savings (line~\ref{alg:fs:break}). If no frequency decrease occurred for the current \gls{pm} ($decrease\_feasible$ stays $False$),\
the procedure will remove \gls{pm}s with higher average $\beta$ and lower electricity price and temperature in line \ref{alg:fs:remove_worse} before continuing.
The idea is that such \gls{pm}s would have even lower energy savings and higher revenue losses, so it is not necessary to consider them.


 \begin{algorithm}[!t]
 \scriptsize{
\caption{Frequency Scaling Stage.}
\label{alg:frequency_scaling}
{
\begin{algorithmic}[1]

\Ensure Reduce CPU frequencies while energy savings exceed revenue losses.
\Procedure{Frequency Scaling Stage}{}

\State $decrease\_feasible \gets False$
\State reset frequency of $\forall pm \in active$ to $f_{max}$\
    \label{alg:fs:reset}

\For {$pm \in active$} \label{alg:fs:pmiterate} 
    \State $f \gets f_{max}$ \
        \Comment{Start the loop at max frequency}
    \State $revenue\_cur \gets get\_revenue(pm, f_{to\_apply})$ \Comment Service revenue, $\forall vm \in pm$
    \State $en\_cost\_cur \gets get\_en\_cost(pm, f_{to\_apply})$\Comment Energy cost of the $pm$
    
    \While{$f>f_{min}$} 
    
        \State $f \gets f-f_{step}$ \
            \label{alg:fs:freqStep}
        \State $revenue\_new \gets get\_revenue(pm, f)$ \
            \Comment {Revenue for the new frequency} \
            \label{alg:fs:get_revenue}
        \State $en\_cost\_new \gets get\_en\_cost(pm, f)$ \
            \Comment {New energy cost} \
            \label{alg:fs:get_cost}
        \State $revenue\_loss \gets revenue\_cur - revenue\_new$ \
        \State $en\_savings \gets en\_cost\_cur - en\_cost\_new$ \
        \If {$en\_savings > revenue\_loss$} \
            \label{alg:fs:comparison}
            \State $revenue\_cur \gets revenue\_new$ \
                \Comment{Update current service revenue}
            \State $en\_cost\_cur \gets en\_cost\_new$ \
                \Comment{Update current energy cost}
            \State $decrease\_feasible \gets True$ \
            \State $f_{to\_apply} \gets f$ \
                \Comment{Update currently selected frequency} \
                \label{alg:fs:apply}
        \Else
            \State break \label{alg:fs:break}
        \EndIf
    \EndWhile
    
    \If {$decrease\_feasible$} \
        \State apply $f_{to\_apply}$ to $pm$
    \Else
        \State remove from $active$: $\forall \hat{pm} \in PMs$ s.t. $\hat{pm}$ has higher $average(\beta_{vm})$ and lower electricity price and temperature than $pm$ \
        \label{alg:fs:remove_worse}
    \EndIf
\EndFor
\EndProcedure

\end{algorithmic}
}
}
\end{algorithm}

\section{Evaluation}
\label{sec:evaluation}

We evaluated the \schedulerfreq{} method in a large-scale simulation of \vmnumsimulation{} \gls{vm}s based on real traces of geotemporal inputs and \gls{vm} CPU-boundedness values. The goal of the evaluation is to show the cost savings attainable using our approach, the impact on service revenue and to analyse the dependence on external factors, such as electricity prices and \gls{vm} workloads.

\subsection{Methodology}

\begin{table*}[!t]
\vspace{-0.3cm}
\centering
\caption{Simulation parameters.}
\label{tab:simulation} 
\vspace{\tablecaptionmargin}

\begin{tabular}{ccccccccccccc}
\hline
 \gls{pm}s & \gls{vm}s & \
 $f_{base}$ & $f_{min}$ & $f_{max}$ & $f_{step}$  & \
 $P_{base}$ & $P_{idle}$ & $P_{dif}$  & \
 $C_{base}$ & $C_{CPU}$ & $C_{RAM}$ & $RAMsize_{base}$  \\ 
 %
 \pmnumsimulation{} & \vmnumsimulation{}& \
 1 GHz & 1.8 GHz & 2.6 GHz & 200 MHz & \ 
150 W & 100 W & 15 W & \ 
 0.027 \$/h  & 0.018 \$/h  & 0.025 \$/h & 1 GB\\ 
\hline
\end{tabular}


\vspace{\tablebottommargin}
\vspace{-3mm}
\end{table*}

The simulations were performed using Philharmonic\footnote{http://philharmonic.github.io/},\
a cloud controller simulator for geographically-distributed clouds\
that we developed \cite{lucanin2014energy}.\ 
A simulation in Philharmonic consists of iterating through the timeline, collecting the currently available electricity prices and temperatures, as well as the incoming \gls{vm} requests. The simulated controller is called to determine cloud control actions, such as \gls{vm} migrations or \gls{pm} frequency scaling. The applied actions are used to compute the resulting energy consumption and electricity costs.

\begin{figure}
\vspace{\figtopmargin}
\vspace{0.1cm}
\centering
\includegraphics[width=1.0\columnwidth]{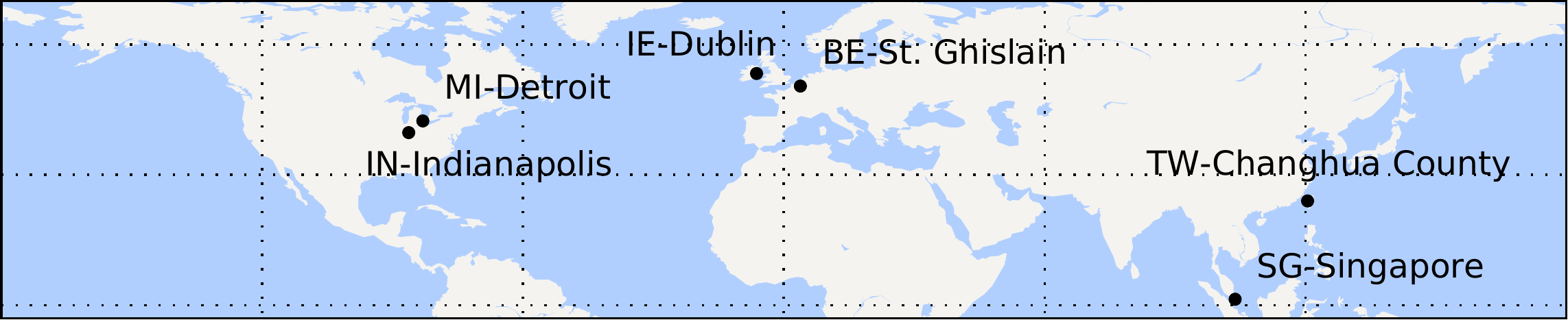}
\vspace{-0.2cm}
\vspace{\figcaptionmargin}
\caption{Cities used as data center locations in the simulation.}
\label{fig:cities}
\vspace{\figbottommargin}
\end{figure}

To compute the energy costs of the simulated geographically-distributed cloud, we consider a use case of six data centers. We used a dataset of real-time electricity prices described in \cite{alfeld_toward_2012}\
and temperatures from the Forecast\footnote{http://forecast.io/} web service.\
The selected data center locations\
(chosen to resemble Google's deployment)\
are shown in Fig.~\ref{fig:cities}. Due to lack of \gls{rtep} data for the four non-US cities, we synthetically generated electricity prices from the data known for other US cities.\
We shifted the time series based on the time zone offsets\
and added a difference in annual mean values to resemble local values.\
Additionally, in the evaluation we analyse a scenario with fixed electricity prices,\
where mean values constant over time are considered for each location.

\begin{figure}[!t]
\vspace{-0.1cm}
\vspace{\figtopmargin}
\centering
\includegraphics[width=0.85\columnwidth]{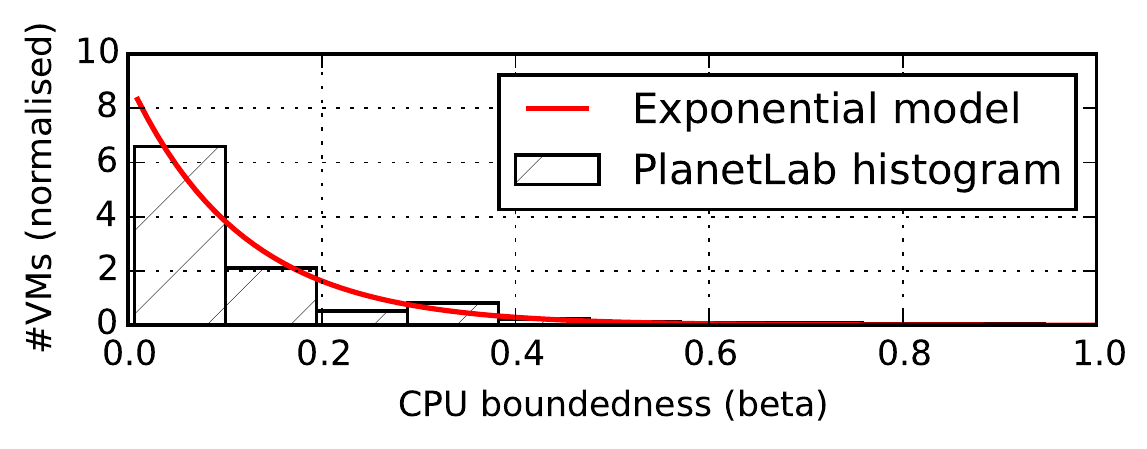}
\vspace{\figcaptionmargin}
\caption{\gls{vm} CPU-boundedness distribution from PlanetLab traces.} 
\label{fig:beta_model}
\vspace{\figbottommargin}
\vspace{-0.3cm}
\end{figure}

To generate realistic \gls{vm} CPU-boundedness values, we analysed the\ 
PlanetLab\footnote{https://github.com/beloglazov/planetlab-workload-traces}\ 
dataset of CPU usage traces for\ 1024 \gls{vm}s 
collected every five minutes throughout a day.\
We mapped each \gls{vm}'s average CPU usage to a $\beta$ value.\ 
From the generated $\beta$ dataset, we fit an exponential distribution, shown in Fig.~\ref{fig:beta_model}.\ 
The empirical histogram of the traces normalised to an area of 1 is also included. We used this model to generate the $\beta$ values of the \gls{vm}s in the simulation. 

We consider the perceived-performance pricing model from\ 
Section~\ref{sec:model}.\ 
We also evaluated the ElasticHosts performance-based pricing model,\
but CPU frequency scaling was not feasible, due to high \gls{vm} prices compared to energy costs.\
Savings are still possible with our perceived-performance pricing model, as for certain less CPU-bound \gls{vm} workloads a CPU frequency reduction does not decrease the service revenue substantially and yet achieves energy savings.

\begin{figure}[!b]
\vspace{\figtopmargin}
\vspace{-0.3cm}
\centering
\includegraphics[width=1.0\columnwidth]{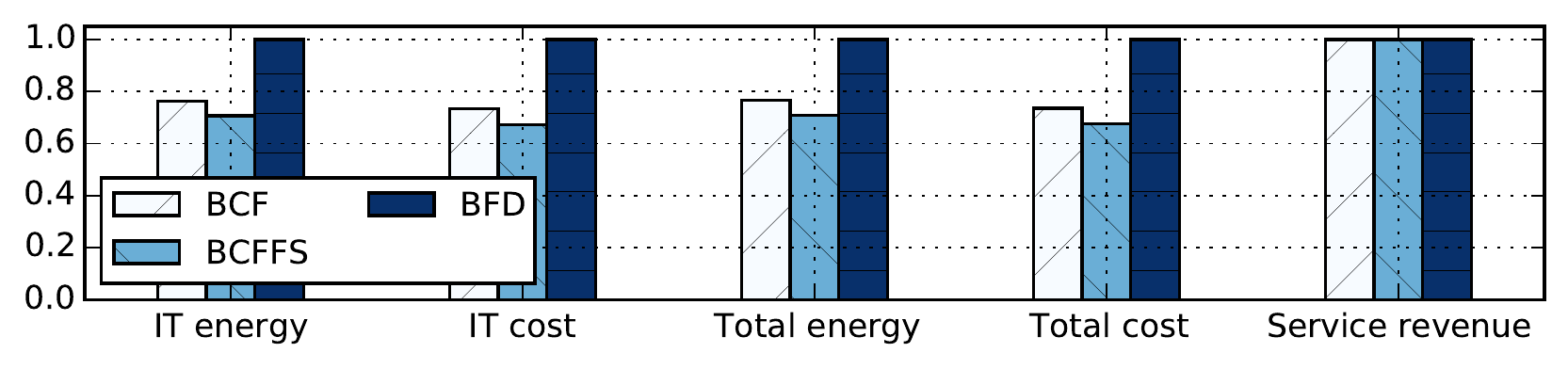}
\vspace{-0.2cm}
\vspace{\figcaptionmargin}
\caption{Aggregated results for a \vmnumsimulation{} \gls{vm} simulation using different controllers.}
\label{fig:aggregated_results-single_simulation}
\vspace{\figbottommargin}
\end{figure}


The parameters used in the simulation are summarised in Table~\ref{tab:simulation}.
To define the cloud,\
the number of total $PMs$ and $VMs$ for the case of \gls{vm} boot requests is given.\
The resource values we assumed in this simulation (number of CPU cores, amount of RAM)\
were uniformly distributed.\
For each \gls{vm}, we used one CPU core and ranged the amount of RAM between 8-32 GB to vary resource\
utilisation over time and \gls{vm} price depending on RAM size.\
\gls{pm}s were varied between 1-4 CPU cores and 16-32 GB RAM, to model a heterogeneous infrastructure.\ 
The boot time and duration of the VM requests were randomly generated following a uniform distribution within the simulation time to vary the duration of the VMs and distribute delete events.\
The total duration of the simulation was set to 168 h (7 days), with 1 h step size.\ 
The simulation step size was selected based on the available geotemporal input datasets, but the cloud controller could be invoked at different periods in production cloud systems\
(e.g. on new \gls{vm} requests or geotemporal input changes).\
Based on the workload and \gls{pm} capacity, at most \maxactivepmnumsimulation{} \gls{pm}s\
were active at once.\ 
Each \gls{pm} can operate in five\
frequency modes\
between a minimum and maximum frequency, $f_{min}$ and $f_{max}$ respectively, in steps of 200 MHz ($f_{step}$), similar to \cite{pierson2011utility}.\
To define the cost models, the parameters used for the power model in Eq.\ref{P_max_f} were based on \cite{pierson2011utility} and the idle power, $P_{idle}$, was assumed to be equal to 50\% of the peak value (at maximum frequency), $P_{peak_{f_{max}}}$, like in \cite{fan2007power}.\
The parameters for the pricing model of Eq. \ref{eq:simple_price} were based on the hourly ElasticHosts~\cite{elastichosts} \gls{vm} prices.\
We also assume the cost of other resources, e.g. disk, that are not used in this study to be fixed.
The specified settings were used in all the experiments, unless otherwise stated (e.g. when certain parameters were varied to measure their impact).


We consider two baseline controllers for results comparison. The first controller is a method for \gls{vm} migration dynamically adapting to user requests using a \schedulerbase{} placement heuristic developed\ 
in \cite{beloglazov_energy-aware_2012}.\
The second baseline controller we developed called \schedulermigr{} is a variant of the \schedulerfreq{} controller that applies \gls{vm} migration based on geotemporal inputs, but no frequency scaling. As the focus of this work is on frequency scaling under performance-based pricing, the \schedulermigr{} baseline method allows us to quantify the improvement brought by this specific aspect.

In the remainder of the section we show the results\ 
for different scenarios to compare the energy savings and revenue loss resulting from applying our cloud controller approach. 

\subsection{Energy Costs and Service Revenue}

\begin{table}[!b]
\vspace{\tabletopmargin}
\vspace{-0.3cm}
\centering
\caption{Absolute aggregated simulation results}
\vspace{\tablecaptionmargin}
\label{tab:aggregated_results} 
\begin{tabular}{lrrr}
\toprule
{} &        BCF &      BCFFS &        BFD \\
\midrule
IT energy (kWh)     &   16226.87 &   15028.00 &   21261.16 \\
IT cost (\$)         &     735.29 &     674.14 &    1002.19 \\
Total energy (kWh)  &   19477.54 &   18043.00 &   25443.02 \\
Total cost (\$)      &     878.80 &     805.86 &    1193.82 \\
Service revenue (\$) &   62995.63 &   62977.79 &   62995.63 \\
\bottomrule
\end{tabular}
\vspace{\tablebottommargin}
\end{table}

We start the analysis of the results\ 
with the aggregated energy costs and service revenue of the proposed method and the considered baseline methods for the \vmnumsimulation{}-\gls{vm} simulation described in Table~\ref{tab:simulation}.\ 
%
%
The aggregated results are shown in Fig.~\ref{fig:aggregated_results-single_simulation}. A column group is shown for each of the examined metrics\ 
-- energy and cost used by the IT equipment, total energy and cost that also include the cooling overhead based on outside temperatures and the service revenue obtained from hosting the \gls{vm}s based on the perceived-performance pricing model. In each group, there is a column for our proposed \schedulerfreq{} controller and the two baseline methods. The values are normalised as a relative value of the \schedulerbase{} baseline controller's results. Absolute values are listed in Table~\ref{tab:aggregated_results}. The proposed \schedulerfreq{} controller achieves \ensavingsmax{} total energy cost savings compared to the \schedulerbase{} controller and \ensavingsfreq{} compared to the \schedulermigr{} controller. No significant service revenue losses are incurred by \schedulerfreq{}.\
Based on our perceived-performance pricing model, this also indicates no significant performance degradation.\
Since similar service revenue and performance results were obtained for other simulations as well, we omit the results for service revenue losses in the rest of the section.

\begin{figure}[!t]
\vspace{\figtopmargin}
\centering
\includegraphics[width=0.9\columnwidth]{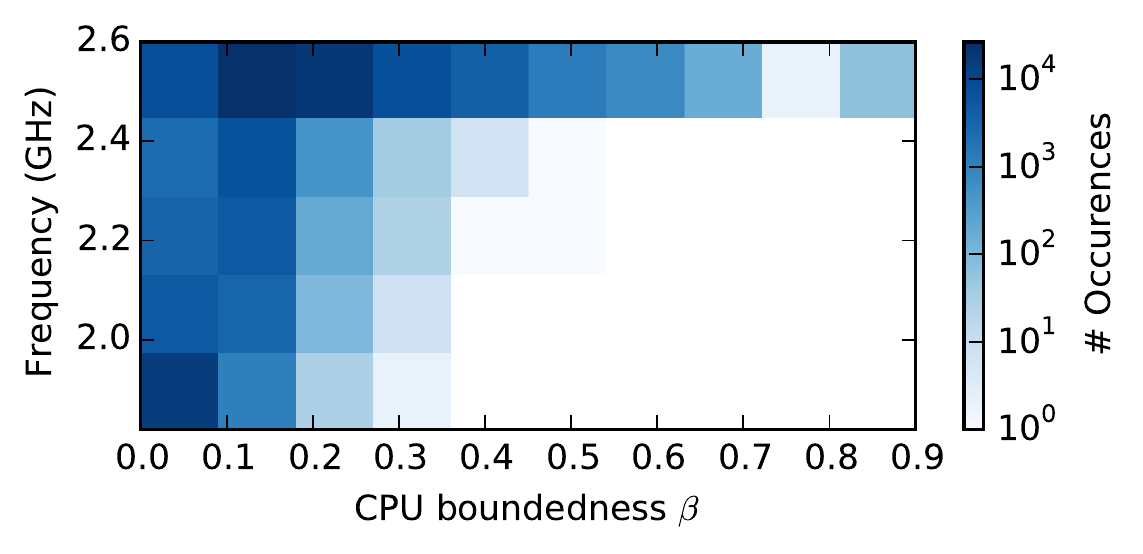}
\vspace{\figcaptionmargin}
\caption{Occurrences of $(\beta, f)$ combinations among the controlled \gls{vm}s.}
\label{fig:beta_freq_histogram}
\vspace{\figbottommargin}
\vspace{-0.2cm}
\end{figure}

To explore the frequencies $f$ assigned to \gls{vm}s in the simulation and compare them with the \gls{vm}s' CPU boundedness $\beta$, we counted the number of occurrences of each $(\beta, f)$ combination for every \gls{vm} and time slot. This data is illustrated\
as a bivariate histogram\
in Fig.~\ref{fig:beta_freq_histogram} with the number of occurrences shown on a logarithmic scale.\
Darker areas show a higher number of frequency occurrences for the respective $(\beta, f)$ combination.\
It can be seen that the occurrences of CPU frequencies assigned based on each \gls{vm}'s CPU boundedness match the areas where \gls{vm} prices are high\ 
based on the pricing model from Fig.~\ref{fig:price_model}.\
The area with high $\beta$ and low $f$, where prices would be the lowest, contains no occurrences.\
The darkest areas of the graph with a high number of occurrences\
represent the balance between energy savings and profit losses,\ 
which is\
in line with the controller requirements\ 
that energy cost savings should be maximised, but not exceeded by revenue losses.

\subsection{Cloud size variation}

\newcommand{\sizevariationpmsmin}{200}
\newcommand{\sizevariationpmsmax}{1.2k}
\newcommand{\sizevariationpmsstep}{200}
\newcommand{\sizevariationvmsmin}{400}
\newcommand{\sizevariationvmsmax}{2.4k}

\begin{figure}[!t]
\vspace{\figtopmargin}
\centering
\includegraphics[width=1.0\columnwidth]{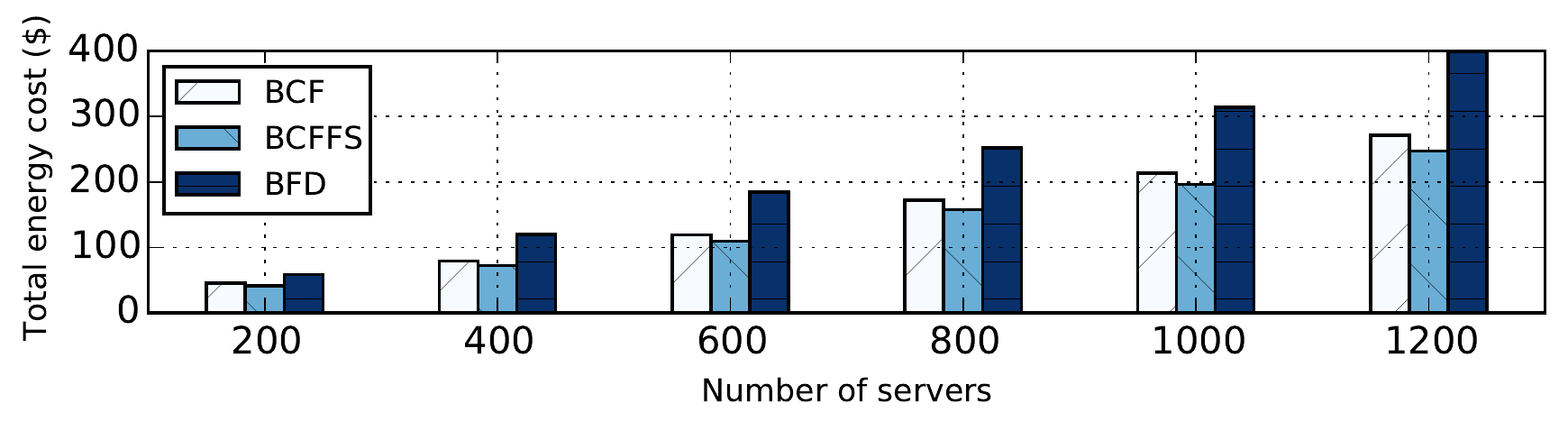}
\vspace{-0.3cm}
\vspace{\figcaptionmargin}
\caption{Energy cost savings for different number of hosts and \gls{vm}s.}
\label{fig:server_num_variation}
\vspace{\figbottommargin}
\end{figure}

We simulated clouds with different numbers of hosts and a proportional frequency of incoming \gls{vm} requests to examine the energy cost savings at different scales. The number of hosts was ranged from \sizevariationpmsmin{} to \sizevariationpmsmax{} with a step size of \sizevariationpmsstep{}. The incoming requests were generated proportionally to the number of hosts to keep the utilisation fixed.\
In Fig.~\ref{fig:server_num_variation}, we show the results in terms of absolute energy costs. We can see that the absolute energy costs savings increase for a larger number of hosts, as the cost of operating the hosts increases. However, the performance of the algorithm is not greatly affected by the number of hosts, achieving similar relative energy cost savings in all the cases.

\begin{figure}[!t]
\vspace{\figtopmargin}
\centering
\includegraphics[width=1.0\columnwidth]{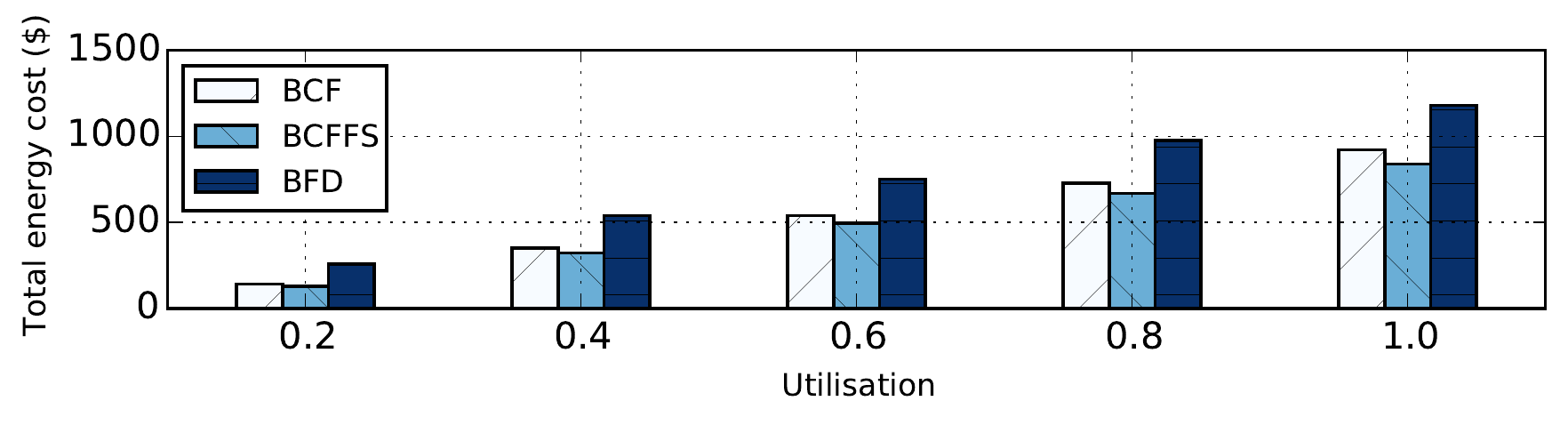}
\vspace{-0.3cm}
\vspace{\figcaptionmargin}
\caption{Energy cost savings for different cloud utilisation rates.}
\label{fig:utilisation_variation}
\vspace{\figbottommargin}
\vspace{-0.2cm}
\end{figure}

\subsection{Utilisation Variation} 

To investigate the performance of the algorithm in terms of energy savings and revenue loss for different utilisation scenarios, we vary the number of \gls{vm} requests for the same number of \gls{pm}s. Energy costs from frequency scaling and revenue loss of the \schedulerfreq{} controller compared to the \schedulermigr{} controller for different utilisation scenarios are shown in Fig.~\ref{fig:utilisation_variation}.\
Although no significant revenue loss occurs for any of the scenarios, energy savings using frequency scaling increase as cloud utilisation increases.\
This is due to the concave shape of the power model, where power decreases faster with CPU frequency reduction for higher utilisation. The \schedulerfreq{} controller is therefore best suited for highly utilised \gls{pm}s.



\subsection{Electricity Cost Variation}



\begin{figure}[!b]
\vspace{\figtopmargin}
\vspace{-0.4cm}
\centering
\includegraphics[width=0.8\columnwidth]{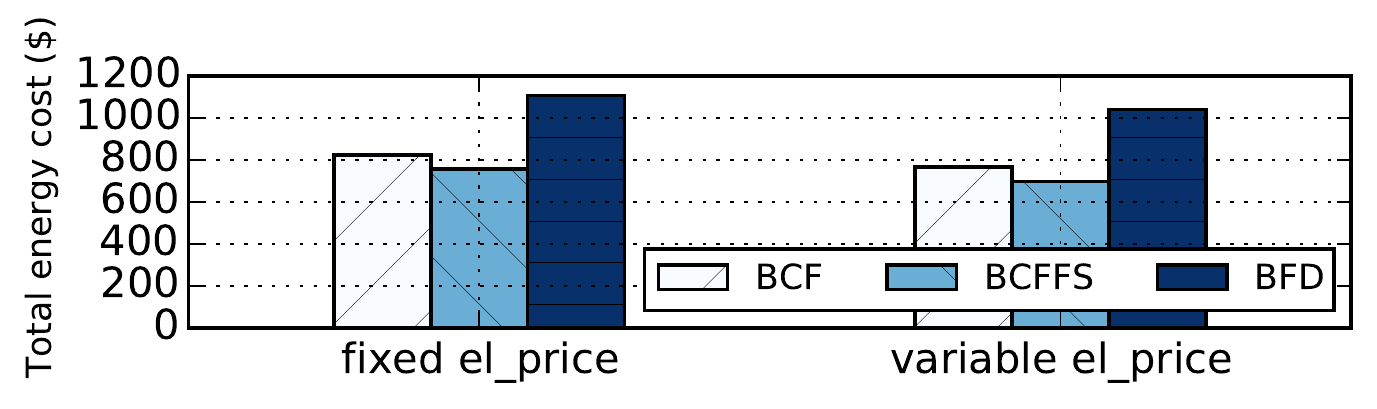}
\vspace{\figcaptionmargin}
\caption{Energy cost savings for fixed and variable electricity pricing options.}
\label{fig:el_price_variation}
\vspace{\figbottommargin}
\end{figure}

Different cloud providers can have access to different electricity pricing schemes. Although some might have access to \gls{rtep}, it is also interesting to see how our controller would perform under fixed electricity pricing.\
In this set of experiments, we compare scenarios for fixed and variable electricity prices to investigate the impact of electricity pricing on the energy savings attainable using the \schedulerfreq{} algorithm.\ 
The results are shown in Fig.~\ref{fig:el_price_variation}. Energy costs are reduced under variable electricity pricing by exploiting runtime information and adapting the cloud configuration within the day according to electricity price changes. In both cases, however, the proposed cloud controller, \schedulerfreq{}, achieves significant savings compared with the baseline algorithms. 

\begin{figure}[!t]
\vspace{\figtopmargin}
\centering
\includegraphics[width=1.0\columnwidth]{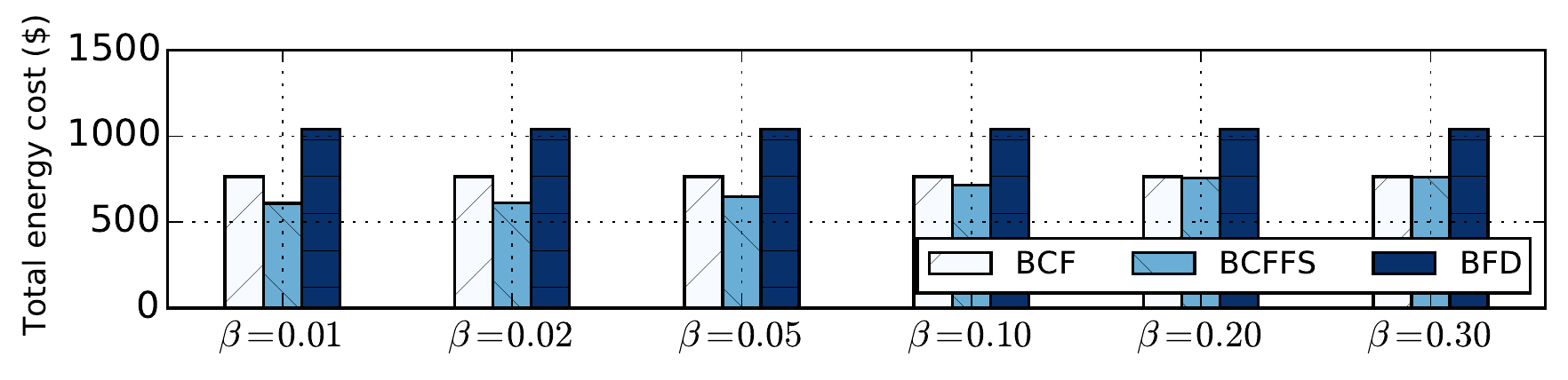}
\vspace{-0.3cm}
\vspace{\figcaptionmargin}
\caption{Energy cost savings for \gls{vm}s with different fixed CPU-boundedness.}
\label{fig:beta_variation}
\vspace{\figbottommargin}
\vspace{-0.1cm}
\end{figure}




\subsection{CPU-Boundedness Variation}

To evaluate the impact of different workloads, we ran\
simulations\
using the same\ 
\gls{pm}s and \gls{vm} requests\
with only the \gls{vm} CPU-boundedness properties\ 
being varied. We generated scenarios with \gls{vm}s of fixed CPU-boundedness properties and evaluated the total energy cost of the cloud controllers for each scenario. The total energy costs\ 
for CPU-boundedness properties ranging from $0.01$ to $0.3$ are shown in Fig.~\ref{fig:beta_variation}.\
With the increase of CPU-boundedness $\beta$, energy costs of
the \schedulerfreq{} controller increase gradually. Between a $\beta$ of $0.05$ and $0.2$ there is a substantial energy cost increase. This happens when the revenue losses exceed energy cost savings from even an initial frequency reduction and at this point no frequency scaling is performed. The \schedulerfreq{} controller achieves the best results for predominantly I/O-bound workloads.


\section{Conclusion}
\label{sec:conclusion}

In this paper we proposed a novel perceived-performance pricing model that would enable applying energy saving actions on \gls{vm}s where CPU frequency scaling would not degrade the performance. We presented a cloud controller that utilises said pricing model and applies \gls{vm} migrations and CPU frequency scaling accounting for the trade-offs of service revenue losses and energy cost savings in a geographically-distributed cloud. We evaluated the proposed cloud controller in a simulation using realistic CPU-boundedness data, electricity prices and temperatures. Our results show significant energy cost savings can be achieved without reducing service revenue. Also, we highlighted parameters that improve the controller's efficiency, such as low workload CPU-boundedness and high \gls{pm} utilisation, that can help cloud providers assess the method's potential applicability.\ 
In the future we plan to\
improve the research with memory power management and a more detailed power model for frequency scaling on multiple cores. This will allow us to more precisely estimate the controller's efficiency for a wider range of \gls{vm} instance types.

\bibliographystyle{IEEEtran-kermit}
\bibliography{references,references-drazen}

\end{document}